\documentclass[11pt]{article}
\usepackage[margin=1in]{geometry}
\usepackage[T1]{fontenc}
\usepackage{lmodern,microtype}
\usepackage{amsmath,amssymb,graphicx,booktabs,array}
\usepackage[colorlinks=true,linkcolor=blue,citecolor=blue,urlcolor=blue]{hyperref}
\hypersetup{pdftitle={Exact Dice Degradation Envelopes for Label-Free Segmentation Update Selection},pdfauthor={Yuvraj Verma},pdfsubject={Segmentation update selection and exact relative Dice degradation bounds},pdfkeywords={segmentation, test-time adaptation, Dice, certification}}
\newtheorem{theorem}{Theorem}
\newtheorem{corollary}{Corollary}
\newcommand{\Dice}{D_{\varepsilon}}
\newcommand{\Cert}{C_{\varepsilon}}
\newcommand{\R}{\mathcal{R}}
\newcommand{\Q}{\mathcal{Q}}
\title{Exact Dice Degradation Envelopes for Label-Free\newline Segmentation Update Selection}
\author{Yuvraj Verma\hspace{0.35em}\href{https://orcid.org/0009-0004-2138-3159}{\raisebox{-1pt}{\includegraphics[height=10pt]{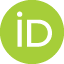}}}\\\texttt{yuvrajverma282004@gmail.com}}
\date{}
\begin{document}
\maketitle
\begin{abstract}
Test-time segmentation updates can improve an unlabeled input while damaging another. Confidence and transformation consistency can rank candidates, but do not in themselves bound the loss caused by returning a changed mask. We derive the exact largest decrease in binary Dice between a baseline and a candidate, maximized over every possible unknown reference mask on the same pixels. A maximizing reference always lies inside baseline foreground: three observable mask counts reduce the problem to at most four reference-cardinality checks. Accepting an output only when its envelope is below a budget gives a deterministic, per-case Dice degradation limit relative to the baseline. A finite grid of logit interpolations permits partial candidate outputs without extra network forwards. Exhaustive and randomized validation checks 5,026,592 reference masks with no discrepancy. We evaluate six independently trained MK-UNet sources, three seeds per task, on ClinicDB, Massachusetts Buildings and the external Kvasir-SEG cohort, including appearance corruptions. Across 17,808 case-condition-seed evaluations, 0 certificate violations occur in the checked candidate comparisons. At a one-point Dice budget, exact versus loose interpolation changes the returned mask in the following fractions of episodes (ClinicDB: 51.0\% versus 38.7\%; Buildings: 62.1\% versus 53.1\%; Kvasir-SEG: 52.3\% versus 34.9\%). Mean Dice effects and useful-update rejection are reported separately from admissibility. An exploratory oracle-presence-bit ablation examines how weak supervision relaxes the empty-baseline restriction. The envelope quantifies relative mask-change risk; it does not certify anatomical correctness, preserve other metrics, or guarantee adaptation benefit.
\end{abstract}

\section{Introduction}
An image-segmentation system may encounter darker illumination, different acquisition characteristics or corrupted inputs after training. Test-time adaptation modifies a model from unlabeled observations, commonly by reducing predictive entropy or changing normalization statistics \cite{tent,eata,sar}. Such updates can improve average overlap while harming individual cases. This matters when model outputs locate a lesion or count small structures: the decision to return a changed mask deserves its own evaluation.

Transformation consistency is an appealing acceptance signal. A prediction that agrees under a horizontal flip looks stable, but two consistently wrong masks can agree. Predicted foreground area is similarly observable, yet an almost unchanged area can conceal a spatial relocation. The question considered here is narrower and checkable: \emph{given the baseline mask and a candidate mask, what is the largest Dice decrease their replacement could cause for any unknown binary reference?}

Even direct Dice agreement between the two outputs does not answer this question. In the unsmoothed case, masks with $a=b=100$ and $c=90$ have pairwise Dice $0.9$. Yet the reference consisting of the ten removed pixels yields a decrement $20/110\approx0.182$, exceeding the pair's Dice distance $0.1$. An exact replacement bound must account for the directional overlap problem, rather than treating output agreement as a metric triangle inequality.

We answer that question exactly. Partitioning the image by membership in the two masks yields four regions. For any fixed reference size, the worst overlap change is a small linear allocation problem with a size-independent region order. We then prove that a maximizing reference never needs pixels outside baseline foreground. Only endpoints and stationary-point neighbors within one interval remain. This gives a certificate computed without labels, assumptions about image distributions, or a learned risk predictor.

The certificate concerns a difference in a specified metric. It allows a poor baseline to remain poor, and an adversarially possible reference can make a genuinely useful update inadmissible. We therefore pair the theorem with an empirical study of its price: useful-update rejection, oracle benefit, foreground-change direction, normalization effects, computation and failures. We also certify partial outputs along a fixed interpolation grid, rather than treating a full candidate and an unchanged baseline as the only options.

The contributions are (i) an exact, attainable all-reference Dice-decrement envelope using three mask counts and at most four cardinalities; (ii) a certificate-based return policy and finite interpolation extension; and (iii) an auditable three-seed study distinguishing the guarantee from achieved adaptation gains, including an explicitly exploratory weak-supervision ablation. Entropy adaptation, rollback and selective routing are established ideas; the claimed object here is the specific mask-pair envelope and its operational tradeoff.

\section{Related work}
\paragraph{Test-time adaptation.} Tent estimates normalization statistics and updates affine parameters through entropy minimization, including a semantic-segmentation setting \cite{tent}. EATA filters unreliable and redundant samples and regularizes important parameters \cite{eata}. SAR addresses instability and sharpness-sensitive updates, including the role of BatchNorm \cite{sar}. CoTTA combines teacher predictions with restoration during continual adaptation \cite{cotta}. Medical segmentation has its own adaptation literature \cite{onthefly}. These methods motivate the candidate updates used here; our gate is not a replacement backbone or a claim to reproduce their classification benchmark scores.

\paragraph{Selective segmentation and risk certification.} HD-TTA generates geometric hypotheses, uses a representation-based selector and skips some confident cases in brain-tumor segmentation \cite{hdtta}. SafeRestore fits selective restoration policies and certifies marginal incident rates on disjoint labeled samples under an image-level i.i.d. working model \cite{saferestore}. These are close conceptual precedents for evaluating a decision layer. Our quantity differs: it is the exact maximum metric decrement for one already observed mask pair, over all possible reference masks. It needs no labeled certification sample and does not estimate a population incident rate. Conversely, it offers no absolute quality or clinical-risk guarantee. Our contribution is scoped to this mathematical object, rather than priority in adaptation, routing or rollback.

\paragraph{Metric-derived bounds.} The Jaccard distance satisfies a triangle inequality, including measure-based variants \cite{jaccard}. Combining that inequality with the Dice--Jaccard transform gives a generic symmetric replacement bound. Our envelope instead maximizes the specified directional decrement exactly and identifies an attaining reference; the distinction is visible even for nested masks. We include the metric-derived rule as a theoretical comparator, rather than treating the availability of any label-free bound as new.

\section{Exact observable Dice envelope}
Let $A,B\subseteq\Omega$ be the baseline and candidate masks on the same $N=|\Omega|$ pixels. Write $a=|A|$, $b=|B|$ and $c=|A\cap B|$. For a reference $Y\subseteq\Omega$, smoothed binary Dice is
\begin{equation}
\Dice(P,Y)=\frac{2|P\cap Y|+\varepsilon}{|P|+|Y|+\varepsilon},\qquad \varepsilon>0.
\label{eq:dice}
\end{equation}
This convention gives one when both masks are empty. Define
\begin{equation}
\Cert(A,B)=\max\left\{0,\max_{Y\subseteq\Omega}[\Dice(A,Y)-\Dice(B,Y)]\right\}.
\label{eq:certificate}
\end{equation}
The maximum is directional and is attainable by a reference mask. It describes potential loss caused by replacing $A$ with $B$, not the probability of that loss.

\subsection{Four-region reduction}
The four regions are removed foreground $\R=A\setminus B$, common foreground $\mathcal C=A\cap B$, background in both $\mathcal O=\Omega\setminus(A\cup B)$, and added foreground $\Q=B\setminus A$. Their capacities are $r=a-c$, $c$, $o=N-a-b+c$ and $q=b-c$. At a fixed reference cardinality $t=|Y|$, the counts of reference pixels in them are $u_R,u_C,u_O,u_Q$, and the decrement is
\begin{equation}
\frac{2(u_R+u_C)+\varepsilon}{a+t+\varepsilon}
-\frac{2(u_C+u_Q)+\varepsilon}{b+t+\varepsilon}.
\label{eq:allocation}
\end{equation}
For this fixed $t$, maximizing the expression is a capacity-constrained linear allocation. Its per-pixel coefficients in the four regions are
\begin{equation}
\frac{2}{a+t+\varepsilon},\quad
\frac{2}{a+t+\varepsilon}-\frac{2}{b+t+\varepsilon},\quad
0,\quad -\frac{2}{b+t+\varepsilon}.
\end{equation}
Thus one greedily fills $R,C,O,Q$ if $b\geq a$, and $R,O,C,Q$ if $b<a$. Ties can be ordered arbitrarily. Crucially, this order is independent of $t$.

\begin{theorem}[Baseline-contained witness and exact finite search]
For every feasible mask pair and $\varepsilon>0$, a reference attaining Eq.~\eqref{eq:certificate} can be chosen inside $A$. Write $r=a-c$. If $b<a$, its size is $r$ and the reference is $A\setminus B$. If $b\geq a$, a maximizing size lies in $[r,a]$, and the envelope requires at most four cardinality checks. Apart from count feasibility, $N$ is unnecessary.
\label{thm:finite}
\end{theorem}
\paragraph{Proof.} Greedy filling gives the exact optimum at every integer $t$: exchanging a selected lower-coefficient pixel for an unselected higher-coefficient pixel cannot decrease the objective. Every capacity-respecting allocation can be realized by a mask. Filling $R$ increases baseline Dice and decreases candidate Dice, so the first segment is increasing. Filling the final region $Q$ does the reverse, so that segment is decreasing.

If $b\geq a$, the order is $R,C,O,Q$. In the $O$ segment, both intersections are fixed at $a,c$, and the derivative is $-(2a+\varepsilon)/(a+t+\varepsilon)^2+(2c+\varepsilon)/(b+t+\varepsilon)^2\leq0$. Therefore a maximum is attained in $[r,a]$, while only baseline pixels are selected.

If $b<a$, the order is $R,O,C,Q$. In the $O$ segment ($t\geq r$), the derivative is $-(2r+\varepsilon)/(a+t+\varepsilon)^2+\varepsilon/(b+t+\varepsilon)^2<0$. To see this, put $d=a-b\leq r$ and $k=b+t\geq r\geq d>0$. The numerator inequality follows from
\begin{equation}
\begin{split}
(2r+\varepsilon)(k+\varepsilon)^2-\varepsilon(k+d+\varepsilon)^2
&\geq d(2k^2+2k\varepsilon-d\varepsilon)>0.
\end{split}
\end{equation}
The next $C$ segment has derivative $(2(a+o)+\varepsilon)/(a+t+\varepsilon)^2-(2(b+r+o)+\varepsilon)/(b+t+\varepsilon)^2<0$, since $b+r=a+q\geq a$ and $a>b$. Hence the maximizing reference is exactly $R$.

For $b\geq a$, throughout the remaining interval $[r,a]$ the intersections are $i_A=t$ and $i_B=t-r$. The objective is
\begin{equation}
F(t)=\frac{2t+\varepsilon}{a+t+\varepsilon}-\frac{2(t-r)+\varepsilon}{b+t+\varepsilon}
=\frac{2qt+2ar+\varepsilon(r+q)}{(a+t+\varepsilon)(b+t+\varepsilon)}.
\label{eq:scalar}
\end{equation}
Its derivative is $K_A/(a+t+\varepsilon)^2-K_B/(b+t+\varepsilon)^2$, where $K_A=2a+\varepsilon$ and $K_B=2(b+r)+\varepsilon$. For $b>a$, the ratio $(a+t+\varepsilon)/(b+t+\varepsilon)$ increases with $t$, so the derivative has at most one zero and switches only from positive to negative. When $\sqrt{K_A}\ne\sqrt{K_B}$, that stationary point is
\begin{equation}
t^*=\frac{\sqrt{K_A}(b+\varepsilon)-\sqrt{K_B}(a+\varepsilon)}{\sqrt{K_B}-\sqrt{K_A}}.
\label{eq:stationary}
\end{equation}
For $b=a$, the objective is decreasing unless the masks are identical, in which case it is zero. Evaluating $r,a$ and, when in the interval, the floor and ceiling of $t^*$ gives at most four integer checks. The witness contains every removed foreground pixel and $t-r$ common foreground pixels, none outside $A$. $\square$

\subsection{Return policy and information limit}
For a chosen Dice-decrement budget $\tau$, return $B$ if $\Cert(A,B)\leq\tau$, otherwise return $A$.
\begin{corollary}[Per-case relative limit]
For every reference mask $Y$, the returned mask $P$ satisfies $\Dice(P,Y)\geq\Dice(A,Y)-\tau$.
\end{corollary}
The statement follows directly from the maximization. It is deterministic for the specified masks, pixel universe, threshold and smoothing constant. It does not assume exchangeability, calibrated confidence, correct source predictions or institution-independent data. It protects Dice only; a boundary score or task-specific clinical endpoint may still deteriorate.

There is also an unavoidable limit. If $A\neq B$, taking $Y=A$ gives $\Dice(A,A)=1>\Dice(B,A)$. A zero-budget universally nondegrading rule must therefore return the identical mask. Nontrivial changes require a positive budget or additional information about plausible references. Restricting the possible reference size or shape might strengthen the envelope, but would introduce assumptions absent from the present guarantee.

The empty-baseline case is especially restrictive. If $A=\varnothing$ and $|B|=b>0$, the exact envelope is $b/(b+\varepsilon)$, attained at an empty reference. With $\varepsilon=10^{-6}$, any introduced foreground exceeds a one-point budget. Thus this guarantee can force persistence of a source model's complete miss, even when the actual reference contains an object. Relative preservation alone cannot resolve that failure.

\begin{corollary}[Composition on one fixed image]
For masks on the same pixel universe,
\begin{equation}
\Cert(A,C)\leq\Cert(A,B)+\Cert(B,C).
\end{equation}
Thus a sequence of returned masks $A_0,\ldots,A_k$ with each $\Cert(A_{j-1},A_j)\leq\tau_j$ has total relative Dice loss at most $\sum_j\tau_j$ for every common reference.
\end{corollary}
For any $Y$, split the score difference through $B$ and bound both terms, then maximize over $Y$. This allows budgets to compose across successive updates or postprocessors on one fixed image. It does not transfer a guarantee between different frames with different references. Likewise $M(A,B)=\max\{\Cert(A,B),\Cert(B,A)\}$ is a genuine metric: triangle inequality follows in both directions, symmetry is explicit, and $Y=A$ establishes positivity for distinct masks. These are mathematical consequences of the envelope, not additional continual-adaptation experiments.

\subsection{What one foreground-presence bit changes}
Suppose an external source supplies whether $Y$ contains foreground. This is additional weak supervision, not the unrestricted label-free setting. Let $\Cert^{+}$ maximize the nonnegative decrement over $|Y|\geq1$, and $\Cert^{-}$ over the single empty reference. For $a>0$, Theorem~\ref{thm:finite} supplies a nonempty maximizing witness for every changed pair, so $\Cert^{+}=\Cert$. For an empty baseline,
\begin{equation}
\Cert^{+}(\varnothing,B)=\max\left(0,\frac{\varepsilon}{1+\varepsilon}-\frac{2\mathbf{1}[b=N]+\varepsilon}{b+1+\varepsilon}\right).
\label{eq:presence}
\end{equation}
The worst allowed reference is a single pixel outside $B$, if one exists, and inside it otherwise. The greedy objective decreases with reference size when $A$ is empty. In either case the envelope is no larger than $\varepsilon/(1+\varepsilon)$, approximately $10^{-6}$ at our smoothing. Thus a correct presence bit removes the near-unit empty-baseline penalty. If the bit says foreground is absent,
\begin{equation}
\Cert^{-}(A,B)=\max\left(0,\frac{\varepsilon}{a+\varepsilon}-\frac{\varepsilon}{b+\varepsilon}\right).
\end{equation}
These guarantees require a correct bit. For example, supplying ``present'' for an actually empty reference can admit a foreground candidate that loses nearly one Dice unit from an empty baseline. Neither the theorem nor the experiments establish a reliable automatic presence detector.

\begin{figure}[t]
\centering\includegraphics[width=\linewidth]{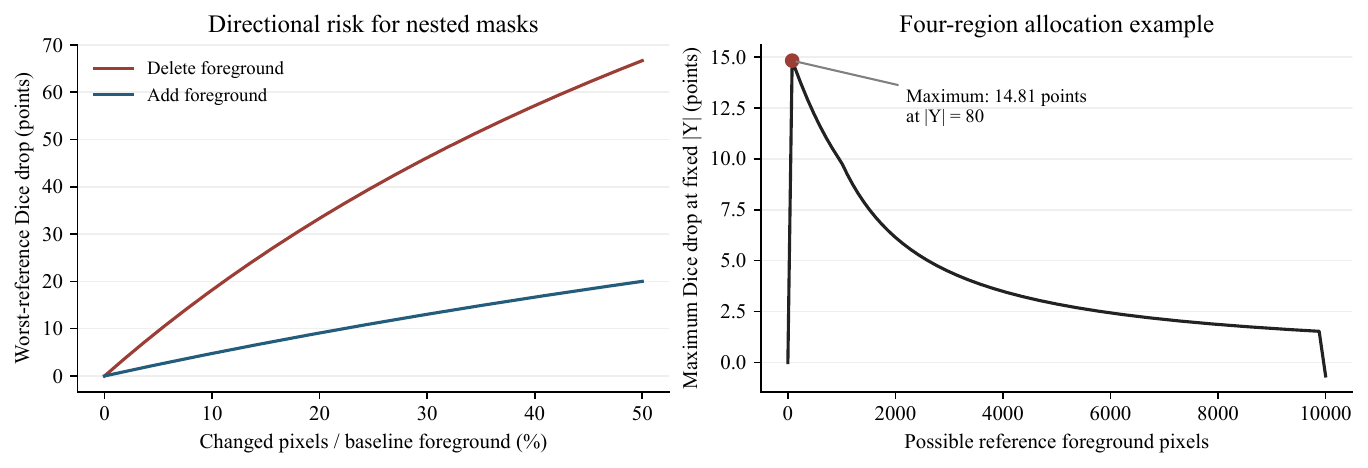}
\caption{Left: exact unsmoothed envelopes for nested, nonempty baseline masks. Removing a fraction $u$ of baseline foreground gives $2u/(1+u)$; adding that fraction gives $u/(2+u)$. Equal numbers of edits can therefore have different directional risk. Right: maximum decrement at each possible reference size for an illustrative $N=10000$, $a=1000$, $b=1040$, $c=920$ mask pair. The finite search recovers the peak without scanning all sizes.}
\label{fig:theory}
\end{figure}

\section{Certified partial outputs}
For frozen and candidate logits $z_0,z_1$, use the fixed mixtures
\begin{equation}
z_\alpha=(1-\alpha)z_0+\alpha z_1,\qquad\alpha\in\{0.125,0.25,0.5,1\}.
\end{equation}
Interpolate these logits to the image's native dimensions and threshold their sigmoid at $0.5$. Certify every resulting binary mask directly, then select the largest admissible $\alpha$; $\alpha=0$ remains available. This policy requires no additional network forwards and makes no monotonicity assumption about the envelope along the grid. Largest $\alpha$ means closest to the full candidate along this prescribed path, not highest unknown accuracy.

For comparison, we use an elementary upper bound
\begin{equation}
U(A,B)=\min\left(1,\frac{2r+\max(b-a,0)}{a}\right),\qquad a>0,
\label{eq:loose}
\end{equation}
with $U=0$ for identical masks and $U=1$ for a changed empty baseline. It follows by expanding the Dice difference, using $i_A-i_B\leq r$ and $2i_B+\varepsilon\leq b+t+\varepsilon$. The exact envelope is no larger. Applying both bounds to the same interpolation grid at the same budget isolates the practical price of a loose bound.

A second, generic bound follows from the metric prior. Add a fixed common atom of measure $\varepsilon/2$ to every mask; the resulting Jaccard distance is
\begin{equation}
d_J(A,B)=\frac{a+b-2c}{a+b-c+\varepsilon/2}.
\end{equation}
Triangle inequality gives $|J(A,Y)-J(B,Y)|\leq d_J(A,B)$. Because $g(x)=2x/(1+x)$ is increasing and concave on $[0,1]$, the largest $g$-difference for arguments separated by at most $d$ is $g(d)-g(0)$. Thus
\begin{equation}
\Cert(A,B)\leq U_J(A,B)=\frac{2d_J(A,B)}{1+d_J(A,B)}.
\label{eq:metricbound}
\end{equation}
This bound is symmetric. For unsmoothed pure addition it gives $2q/(a+2q)$, whereas the exact directional envelope is $q/(2a+q)$. The metric rollback comparator applies $U_J$ to the primary full candidate at the same one-point budget. It was added during theoretical review, using saved observable counts; it does not change the running GPU benchmark or any predeclared update setting.

Equivalently, writing $\delta=1-\Dice(A,B)$ gives $U_J=4\delta/(1+3\delta)$. This supplies a valid generic correction to the uncorrected pairwise-Dice heuristic, while the directional envelope can be substantially smaller.

\begin{figure}[t]
\centering\includegraphics[width=\linewidth]{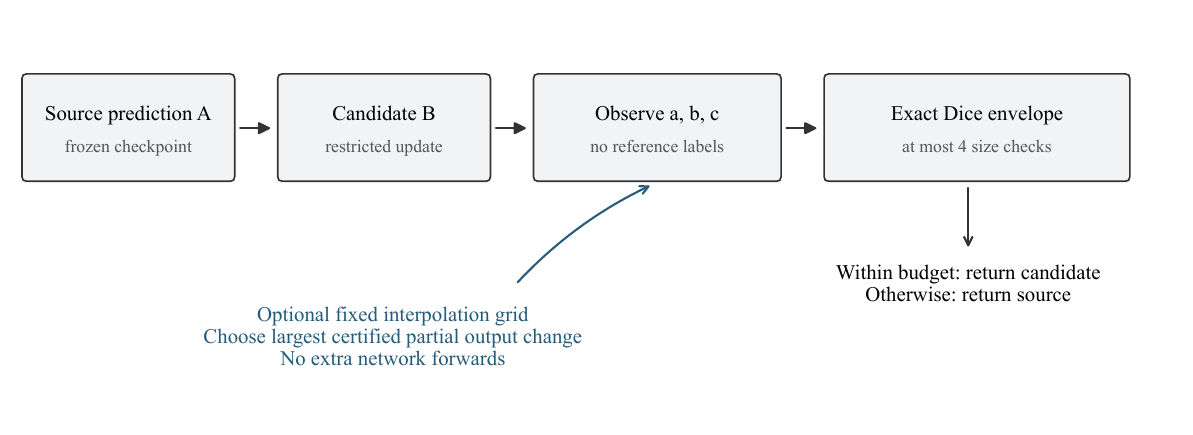}
\caption{The certificate operates on already produced masks. Candidate generation and output selection are separate. The partial-output grid reuses logits; scoring against reference annotations occurs only after every selection is finalized.}
\label{fig:framework}
\end{figure}

\section{Experimental protocol}
\subsection{Sources and data roles}
We use the unmodified 315,566-parameter MK-UNet \cite{mkunet}, at upstream revision \texttt{2fa8b230a057} (full revision recorded in the release manifest). Separate polyp and building models are trained for seeds 101, 102 and 103 on an RTX 4050 laptop GPU with 6GB memory. Each uses 120 epochs, batch 16, AdamW learning rate and weight decay $10^{-4}$, the upstream structure loss, elementwise gradient clipping to $[-0.5,0.5]$, mixed precision, one input scale, and no augmentation or scheduler. The highest raw native-validation mean Dice selects the checkpoint.

ClinicDB \cite{clinicdb} uses the MK-UNet authors' 489/61/62 image split. The reused loader PIL-resizes RGB inputs and masks to 352, then torch-resizes training and inference inputs to 256 (bilinear, \texttt{align\_corners=False}) and training masks to 256 (nearest). Massachusetts Buildings uses the official 137/4/10 tile split \cite{mnih}; a 1500-pixel tile is resized to 512 and divided into four 256-pixel patches. All patches of a parent tile stay in one partition. Native masks are retained for validation and testing. Predictions are stitched before resizing to a tile's native dimensions.

Kvasir-SEG \cite{kvasir} provides 1000 external polyp image--mask pairs. ClinicDB-trained sources are evaluated without using Kvasir labels to train, choose checkpoints or set gates. Audits found zero exact file/decoded-image duplicates between Kvasir and any ClinicDB partition. Exact decoded hashes also separate the internal partitions. These checks do not establish patient, video or spatial independence: such group identifiers are not available in the used manifests.

The earlier screening evaluated these internal test cohorts with a different 200-epoch, multiscale recipe. Those results motivated this extension and are not reported as new source results. The internal cohorts have therefore been previously inspected. The new settings were fixed before external outcomes; the complete protocol, implementation clarifications, fixed interpolation extension and hashes accompany the study.

\begin{table}[t]
\centering\small
\caption{New source runs. Test Dice and IoU are percentages at native resolution. Each row is an independently trained checkpoint selected only on validation. Source evaluation uses batches of up to 16; adaptation evaluation uses single-image episodes (four patches for a building tile), with resize and normalization applied in an algebraically equivalent order. Small numerical differences between the two frozen evaluations are retained rather than equating their scores.}
\label{tab:sources}
\begin{tabular}{llrrrr}
\toprule Domain & Seed & Selected epoch & Val. Dice & Test Dice & Test IoU\\\midrule
ClinicDB & 101 & 111 & 81.35 & 80.47 & 71.57 \\
ClinicDB & 102 & 49 & 81.27 & 81.36 & 73.81 \\
ClinicDB & 103 & 113 & 81.22 & 84.78 & 76.85 \\
Buildings & 101 & 57 & 68.97 & 71.15 & 55.34 \\
Buildings & 102 & 77 & 68.89 & 71.51 & 55.75 \\
Buildings & 103 & 113 & 69.78 & 71.70 & 56.04 \\
\bottomrule
\end{tabular}
\end{table}

\subsection{Candidate generation and controls}
The primary candidate updates only 10,872 BatchNorm affine parameters while keeping source running statistics frozen. It minimizes binary entropy balanced between baseline-predicted foreground and background for one SGD step at learning rate $0.01$. We predeclare one-step $0.001$ and three-step $0.01$ variants. Each episode starts from the selected source checkpoint.

Controls are frozen inference, horizontal-flip probability averaging, a normalization-only prediction, ungated updates, and the earlier flip-Jensen--Shannon/area proxy. The proxy accepts the primary candidate if flip disagreement decreases by at least $10^{-6}$ and predicted input-resolution foreground fraction changes by at most $0.05$. Exact gates use budgets $0.001,0.005,0.01,0.02,0.05$; one Dice point ($0.01$) is the primary illustrative limit, not a clinical tolerance. The loose-bound gate and both interpolation policies use that same limit.

We also implement per-pixel binary Tent-seg, EATA-style and SAR-style candidates with recomputed BatchNorm statistics. Tent-seg uses one entropy step at $0.001$. EATA-style uses three steps, reliable-pixel entropy threshold $0.4\log 2$, nonredundant cosine similarity below $0.95$, entropy weighting and a Fisher penalty of 2000. Fisher is estimated from the first 32 source-training inputs using frozen pseudo labels. SAR-style uses two reliable-entropy gradient evaluations in one SAM update, radius $0.05$, and learning rate $0.001$. The original continual entropy-EMA recovery is omitted because every episode resets. These are transparent binary segmentation adaptations, not exact author classification experiments. We evaluate their ungated outputs and the same exact one-point gate.

\subsection{Shifts, scoring and uncertainty}
Internal tests include clean inputs and three severities each of brightness (0.8/0.65/0.4), gamma (1.3/1.8/2.5), Gaussian noise SD (0.03/0.08/0.15 on $[0,1]$), and Gaussian blur sigma (0.8/1.5/2.5). Noise seeds are fixed per dataset, image and condition, shared across training seeds. The external cohort uses clean and the four middle severities. These image manipulations are controlled stress tests, not evidence of realistic scanner failures. External clean Kvasir represents a real dataset shift, subject to the source/target independence limitations above.

Corruptions operate after input resizing, on 256-pixel inputs or each 256-pixel building patch; blur uses reflection at patch borders. They therefore do not correspond to a common physical scale across datasets. Evaluation and certificates still operate on the returned native masks.

Original label-free decisions use native binary masks on dimensions read from the input image. Only after all candidate masks and those gate decisions are finalized is the reference annotation read for scoring. Dice uses $\varepsilon=10^{-6}$; IoU uses the same stabilizer. Boundary F1 uses the inner eight-connected boundary and a two-pixel Chebyshev tolerance at native resolution, not physical distance. The later exploratory oracle-presence-bit analysis deliberately relaxes the information setting: it derives the single foreground-present/absent bit from the annotation and applies Eqs.~\eqref{eq:presence} and its absent-reference counterpart to saved candidate counts. It does not change the models, candidates or primary label-free outcomes.

We report mean metrics, accepted counts, Dice losses exceeding one point, worst decrement, beneficial rejection (primary candidate gain $>0.001$), and fraction of available positive primary-candidate gain retained. A labeled oracle returns the primary candidate only when its measured Dice improves; it is a diagnostic upper bound and never a deployed policy. Confidence intervals use 2000 bootstrap draws of image/tile identities, averaging all corresponding seeds and conditions jointly. They are conditional case-resampling intervals; repeated corruptions and seeds are not independent patients. Seed standard deviations are reported separately. GPU timings synchronize and warm every update path; data reading and labeled scoring are excluded from method timings and retained in total-case records.

\section{Results}
\subsection{Validation of the envelope}
Alongside the unrestricted envelope, the same reference enumeration independently tests both foreground-presence conditions against their restricted reference sets, including high-smoothing cases and the full-candidate edge case in Eq.~\eqref{eq:presence}.
The sharper implementation matches independent brute-force evaluation for every baseline/candidate/reference triple on one to five pixels, plus 150 random mask pairs at each size six to twelve with every reference enumerated. Four smoothing constants ($10^{-6},0.1,1,100$) yield 5,026,592 reference comparisons and maximum absolute discrepancy 0. Another 10,000 random feasible count configurations, with pixel universes up to $10^8$ and smoothing spanning $10^{-8}$ to $10^8$, match the general four-region kernel. Empty, identical, nested and disjoint masks are included. Floating-point validation is not a formal proof of machine arithmetic; Theorem~\ref{thm:finite} establishes the mathematical result.

\begin{figure}[t]
\centering\includegraphics[width=.78\linewidth]{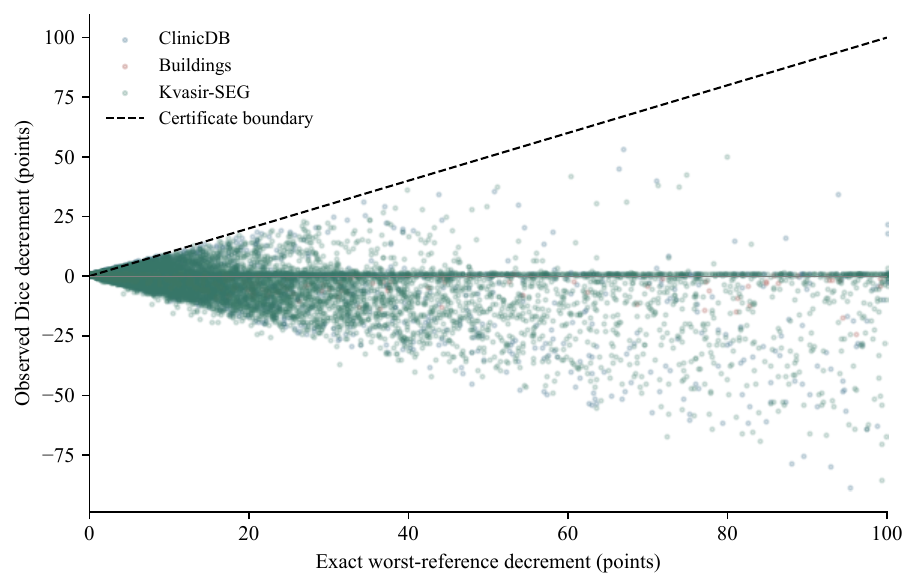}
\caption{Exact certificate versus observed Dice decrement for the primary full candidates. Points below zero are actual improvements. All possible references are bounded by the dashed line, whereas large actual gains can still have large worst-reference envelopes. The repeated image/condition/seed points do not constitute independent clinical observations.}
\label{fig:observed}
\end{figure}

In the real-image benchmark, every checked candidate's observed decrement is at most its certificate (numerical tolerance $10^{-10}$). The independent record audit recomputes 142,464 candidate-versus-baseline certificates and 195,888 saved metric values across all 17,808 episodes, finding no bound violation or metric-arithmetic discrepancy. This observation is an implementation check; it is not the basis of the all-reference theorem. Figure~\ref{fig:observed} shows why certificate and measured harm should not be conflated: useful candidates can lie well below the worst-reference boundary.

\begin{table}[t]
\centering\scriptsize
\caption{Primary policy comparison, uniformly averaging the predeclared conditions within each cohort and the three training seeds. Dice is mean $\pm$ training-seed SD; changes are Dice points relative to the same frozen source. Harm counts case-condition-seed losses $>1$ point; these are not independent cases. Acc. counts admitted candidates or nonzero alpha; Changed counts actual binary-mask differences. Metric rollback is the added theoretical comparator in Eq.~\eqref{eq:metricbound}.}
\label{tab:main}
\begin{tabular}{llrrrrr}
\toprule Cohort & Policy & Dice & Change & Harm & Acc. (\%) & Changed (\%)\\\midrule
ClinicDB & Frozen & $50.15\pm1.47$ & +0.000 & 0 & 0.00 & 0.00 \\
 & Full update & $52.62\pm1.34$ & +2.470 & 507 & 100.00 & 88.71 \\
 & JS/area proxy & $50.54\pm1.58$ & +0.387 & 98 & 25.85 & 21.88 \\
 & Exact rollback & $50.15\pm1.48$ & -0.001 & 0 & 28.62 & 17.33 \\
 & Loose rollback & $50.15\pm1.48$ & +0.001 & 0 & 20.72 & 9.43 \\
 & Metric rollback & $50.15\pm1.47$ & +0.001 & 0 & 16.25 & 4.96 \\
 & Exact interpolation & $50.13\pm1.47$ & -0.019 & 0 & 62.37 & 50.95 \\
 & Loose interpolation & $50.14\pm1.48$ & -0.007 & 0 & 50.08 & 38.67 \\
\midrule
Buildings & Frozen & $35.04\pm6.38$ & +0.000 & 0 & 0.00 & 0.00 \\
 & Full update & $36.63\pm6.33$ & +1.592 & 0 & 100.00 & 91.03 \\
 & JS/area proxy & $35.08\pm6.33$ & +0.047 & 0 & 4.36 & 3.85 \\
 & Exact rollback & $35.04\pm6.37$ & +0.006 & 0 & 38.21 & 29.23 \\
 & Loose rollback & $35.04\pm6.37$ & -0.001 & 0 & 32.05 & 23.08 \\
 & Metric rollback & $35.04\pm6.38$ & -0.001 & 0 & 16.92 & 7.95 \\
 & Exact interpolation & $35.08\pm6.37$ & +0.048 & 0 & 71.03 & 62.05 \\
 & Loose interpolation & $35.05\pm6.37$ & +0.018 & 0 & 62.05 & 53.08 \\
\midrule
Kvasir-SEG & Frozen & $34.32\pm5.37$ & +0.000 & 0 & 0.00 & 0.00 \\
 & Full update & $36.82\pm5.10$ & +2.502 & 2224 & 100.00 & 95.76 \\
 & JS/area proxy & $34.67\pm5.38$ & +0.356 & 237 & 21.15 & 19.43 \\
 & Exact rollback & $34.32\pm5.37$ & +0.003 & 0 & 13.96 & 9.72 \\
 & Loose rollback & $34.32\pm5.37$ & +0.001 & 0 & 9.37 & 5.13 \\
 & Metric rollback & $34.32\pm5.37$ & +0.001 & 0 & 7.02 & 2.78 \\
 & Exact interpolation & $34.35\pm5.35$ & +0.034 & 0 & 56.60 & 52.29 \\
 & Loose interpolation & $34.33\pm5.37$ & +0.010 & 0 & 39.17 & 34.86 \\
\bottomrule
\end{tabular}
\end{table}

\subsection{Useful adaptation versus bounded changes}
On ClinicDB, frozen mean Dice is 50.15\%. The ungated primary update changes it by +2.470 points and causes 507 losses exceeding one point. Exact rollback accepts 28.62\% of full candidates and changes mean Dice by -0.001 points. Exact interpolation returns a nonzero alpha in 62.37\% of episodes; the binary mask actually changes in 50.95\%, versus 38.67\% for loose interpolation. Its mean Dice change is -0.019 points (conditional 95\% interval [-0.047, +0.007]).

On Buildings, frozen mean Dice is 35.04\%. The ungated primary update changes it by +1.592 points and causes 0 losses exceeding one point. Exact rollback accepts 38.21\% of full candidates and changes mean Dice by +0.006 points. Exact interpolation returns a nonzero alpha in 71.03\% of episodes; the binary mask actually changes in 62.05\%, versus 53.08\% for loose interpolation. Its mean Dice change is +0.048 points (conditional 95\% interval [+0.035, +0.058]).

On Kvasir-SEG, frozen mean Dice is 34.32\%. The ungated primary update changes it by +2.502 points and causes 2224 losses exceeding one point. Exact rollback accepts 13.96\% of full candidates and changes mean Dice by +0.003 points. Exact interpolation returns a nonzero alpha in 56.60\% of episodes; the binary mask actually changes in 52.29\%, versus 34.86\% for loose interpolation. Its mean Dice change is +0.034 points (conditional 95\% interval [+0.027, +0.042]).

Exact interpolation and the exact-gated candidates satisfy the one-point loss limit in every observed episode. Their average gains can be small despite this limit. Greater admitted interpolation does not guarantee greater true Dice; the bound uses the same worst-reference criterion independently of observed labels.

\begin{figure}[t]
\centering\includegraphics[width=\linewidth]{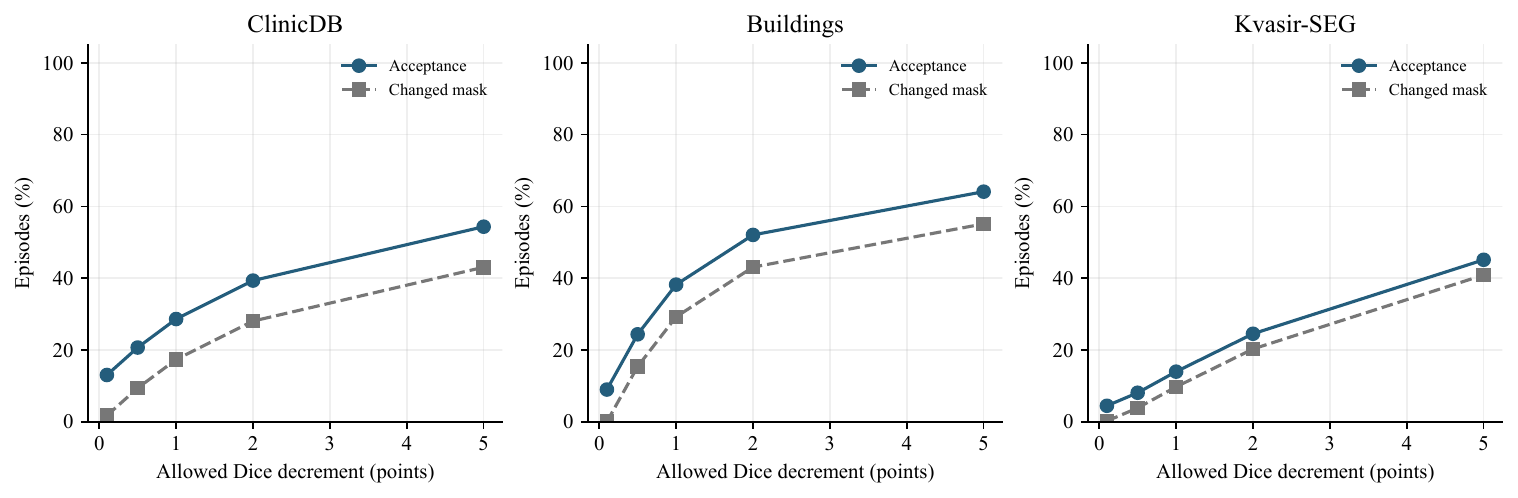}
\caption{Admissibility and actual binary-mask changes for the primary full candidate at every predeclared Dice-decrement budget. Admitting an unchanged mask is distinguished from adapting a returned segmentation. Conservative guarantees can incur large rejection costs. Curves average the fixed conditions and seeds; mixture definitions differ between internal and external cohorts.}
\label{fig:coverage}
\end{figure}

\begin{figure}[t]
\centering\includegraphics[width=\linewidth]{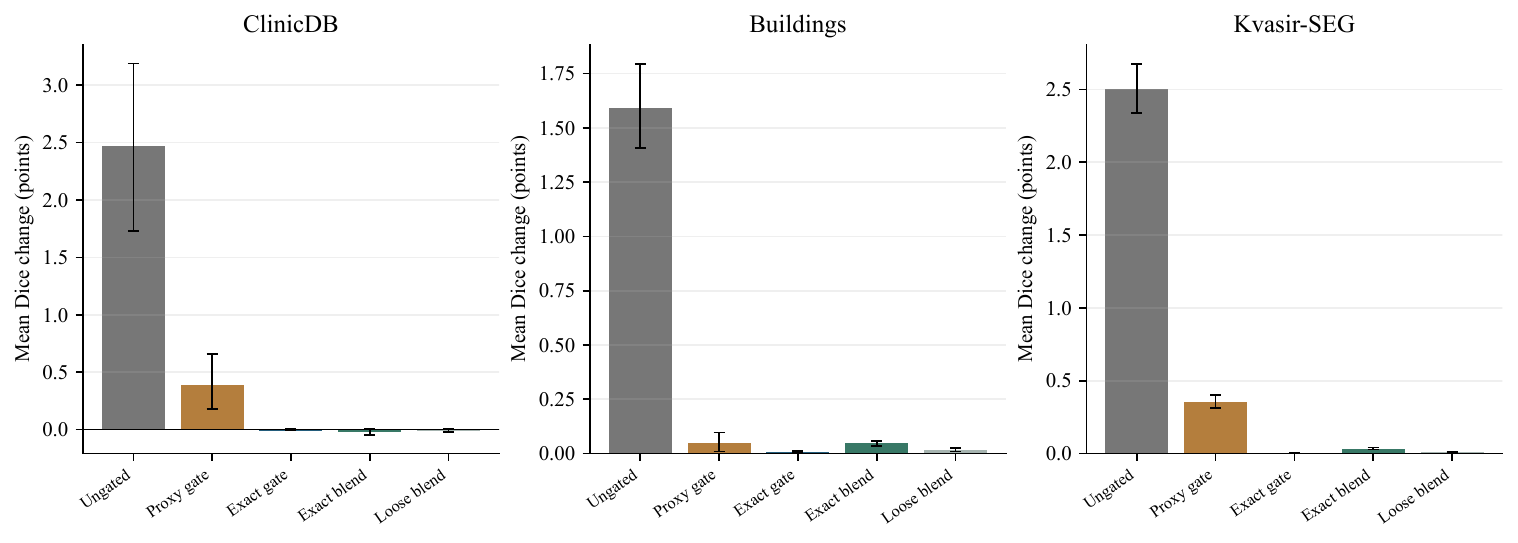}
\caption{Mean Dice changes and conditional paired case-bootstrap intervals. Exact and loose interpolation policies choose from the same fixed alpha grid and use the same one-point risk limit. No test-best operating point is selected.}
\label{fig:tradeoff}
\end{figure}

The one-point certificate must bound a possible decline for any reference, even when the actual reference makes the candidate beneficial. Rejection therefore cannot be interpreted as evidence that an update was incorrect. Figure~\ref{fig:coverage} reports all predeclared budgets; we do not select the most favorable test budget. Figure~\ref{fig:tradeoff} separates full updates, the proxy, exact rollback, and the two interpolation policies. Their conditional intervals quantify case variation under the stated resampling assumptions.

Empty frozen predictions occur in 278 ClinicDB episodes, 35 Buildings episodes, 644 Kvasir-SEG episodes. Across the six update variants, at least one nonempty candidate is proposed in 278, 35, 644 of those episodes, respectively; every such output exceeds the unrestricted one-point budget. In 255, 35, 605 of those episodes at least one proposed recovery actually improves Dice. These repeated observations illustrate the information limit, rather than an implementation failure.

The primary-candidate oracle gains +3.657 points on ClinicDB. Exact full-candidate rollback retains 0.51\% of the available positive primary gain and rejects 812 candidates whose measured gain exceeds 0.1 point. The primary-candidate oracle gains +1.611 points on Buildings. Exact full-candidate rollback retains 0.78\% of the available positive primary gain and rejects 199 candidates whose measured gain exceeds 0.1 point. The primary-candidate oracle gains +3.152 points on Kvasir-SEG. Exact full-candidate rollback retains 0.34\% of the available positive primary gain and rejects 7770 candidates whose measured gain exceeds 0.1 point. The diagnostic oracle is unavailable at inference. Interpolated masks form a different candidate set, so their positive gain is not described as a fraction retained from the full-candidate oracle.


\subsection{Reference candidates and clean-data behavior}
For clean ClinicDB, the primary full update changes mean Dice by +1.266 points, exact rollback by -0.014, and exact interpolation by -0.043. Over the full fixed mixture, normalization alone changes Dice by +10.518 points. For clean Buildings, the primary full update changes mean Dice by -0.001 points, exact rollback by -0.011, and exact interpolation by -0.006. Over the full fixed mixture, normalization alone changes Dice by +18.600 points. For clean Kvasir-SEG, the primary full update changes mean Dice by +1.370 points, exact rollback by +0.002, and exact interpolation by +0.031. Over the full fixed mixture, normalization alone changes Dice by +7.505 points.

Changing normalization statistics can move a segmentation considerably before an optimizer acts. The normalization-only comparator therefore matters when interpreting Tent-seg, EATA-style and SAR-style. The exact gate is candidate-independent, but an uninformative or extensively changed candidate can be rejected on almost every image. A guarantee achieved through rollback should be reported together with how often meaningful changes survive.

\begin{table}[t]
\centering\scriptsize
\caption{Additional candidate comparisons. Mean Dice change (points) and accepted fractions are shown for the ungated reference variants and their exact one-point gates. These binary/episodic adaptations differ from published classification implementations.}
\label{tab:references}
\begin{tabular}{llrrr}
\toprule Cohort & Candidate & Ungated change & Gated change & Accepted (\%)\\\midrule
ClinicDB & Tent-seg & +10.525 & +0.000 & 0.00 \\
 & EATA-style & +10.479 & +0.000 & 0.00 \\
 & SAR-style & +10.520 & +0.000 & 0.00 \\
Buildings & Tent-seg & +18.599 & +0.000 & 0.00 \\
 & EATA-style & +18.603 & +0.000 & 0.00 \\
 & SAR-style & +18.600 & +0.000 & 0.00 \\
Kvasir-SEG & Tent-seg & +7.507 & +0.000 & 0.00 \\
 & EATA-style & +7.554 & +0.000 & 0.00 \\
 & SAR-style & +7.506 & +0.000 & 0.00 \\
\bottomrule
\end{tabular}
\end{table}

\subsection{Exploratory oracle-presence-bit ablation}
The oracle bit indicates nonempty foreground in 17,808 of 17,808 episodes and absence in 0. With this bit, the primary full-candidate gate changes mean Dice by +0.002 points on ClinicDB, +0.006 points on Buildings, +0.003 points on Kvasir-SEG. Table~\ref{tab:presence} includes the same analysis for all three reference variants. Since every reference is nonempty, differences from the unrestricted gate arise only in empty-baseline episodes, as predicted by Eq.~\eqref{eq:presence}.
This analysis was added during theoretical review after the benchmark began; it is not a predeclared primary comparison. The one-point budget and all original candidate updates remain fixed. An oracle bit isolates the information requirement; it is not an operationally available label-free selector. The annotations are not used to optimize a candidate or choose its best setting, but their foreground-presence bit is intentionally supplied to this additional gate. The cohorts' observed presence balance and the absence of a validated presence detector constrain any deployment interpretation.

\begin{table}[t]
\centering\scriptsize
\caption{Exploratory weak-supervision ablation. Mean Dice changes in points for the same full candidates under an unrestricted exact gate and a gate given the oracle foreground-presence bit. Changed percentages count actual returned binary-mask differences. This table must not be interpreted as a label-free performance claim.}
\label{tab:presence}
\begin{tabular}{llrrrr}
\toprule Cohort & Candidate & Universal change & One-bit change & Universal changed (\%) & One-bit changed (\%)\\\midrule
ClinicDB & Full update & -0.001 & +0.002 & 17.33 & 17.54 \\
 & Tent-seg & +0.000 & +5.867 & 0.00 & 11.50 \\
 & EATA-style & +0.000 & +5.863 & 0.00 & 11.50 \\
 & SAR-style & +0.000 & +5.866 & 0.00 & 11.50 \\
Buildings & Full update & +0.006 & +0.006 & 29.23 & 29.23 \\
 & Tent-seg & +0.000 & +4.694 & 0.00 & 8.97 \\
 & EATA-style & +0.000 & +4.694 & 0.00 & 8.97 \\
 & SAR-style & +0.000 & +4.694 & 0.00 & 8.97 \\
Kvasir-SEG & Full update & +0.003 & +0.003 & 9.72 & 9.77 \\
 & Tent-seg & +0.000 & +1.538 & 0.00 & 4.29 \\
 & EATA-style & +0.000 & +1.542 & 0.00 & 4.29 \\
 & SAR-style & +0.000 & +1.538 & 0.00 & 4.29 \\
\bottomrule
\end{tabular}
\end{table}

\subsection{Computation and illustrative cases}
Six source runs contain 720 completed epochs and take 35.0 minutes in total, with peak allocated training memory 2.80 GiB. Dataset/method median certificate arithmetic times in the running benchmark range from 26.0 to 32.9 microseconds; this excludes native mask interpolation and counting. Its largest realized cardinality search has 7 checks. A separate CPU batch benchmark of the saved count triples gives 1.54--1.63 microseconds per call for the sharper routine versus 5.62--6.59 for the general routine (median of five alternating-order repetitions of mean call time). These CPU blocks exclude mask processing, GPU work and disk reads. Evaluation peak allocated memory reaches 0.59 GiB. Table~\ref{tab:timing} reports synchronized operation costs, including the additional work needed for partial-output selection.
The envelope's arithmetic is constant-size after mask counts. Pixel reading and reduction remain $O(N)$ and can dominate it at high resolution; interpolation has additional pixel costs. Source training and candidate generation are therefore not made free by the theorem. The GPU benchmark retains the original general four-region kernel, which checks at most thirteen sizes. The sharper four-check reduction was derived during proof review and is compared against that kernel on every saved mask pair; its CPU timing is reported separately. Measurement details distinguish synchronized candidate operations, native-mask certificate overhead, additional proxy forwards, data reading and offline scoring. Peak memory includes the evaluation implementation and native scoring tensors, rather than an architecture-only inference footprint.

\begin{figure}[t]
\centering\includegraphics[width=\linewidth]{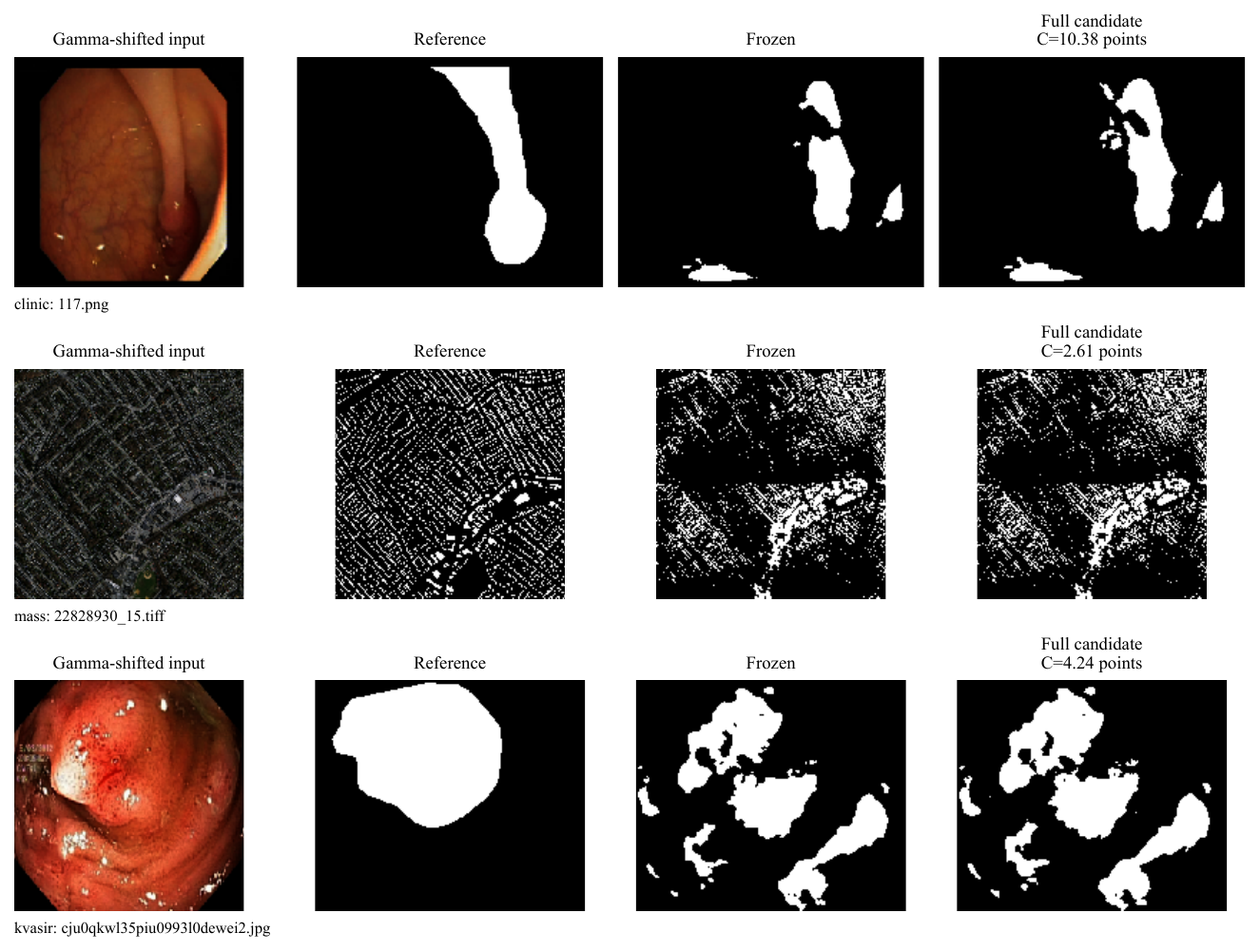}
\caption{Predeclared illustrative cases from ClinicDB, Massachusetts Buildings and Kvasir-SEG \cite{clinicdb,mnih,kvasir}, seed 101, gamma 1.8. The input is displayed at the model's processed resolution; references and predictions are native masks scaled for presentation. Full candidates are shown to explain the certificate's input, not to imply they were returned by the one-point policy.}
\label{fig:qualitative}
\end{figure}

Figure~\ref{fig:qualitative} uses the first test image/tile of each manifest and seed 101 under gamma 1.8, fixed before inspecting these outcomes. The masks illustrate changed support rather than diagnose why an error occurred. Their full-candidate certificates are relative to the displayed frozen masks, and the actual gate can return a smaller partial change or roll back entirely.


\section{Discussion and limitations}
An exact bound answers what could happen under the weakest assumptions. It cannot reveal which reference is actually present. Its tightness means no smaller universal bound can be obtained from the same observed mask pair; a reference realizing the maximum exists. Operationally, this explains why a certifier can reject a beneficial update without being mathematically loose. Assumptions about anatomical shape, foreground size or local reference support could reduce pessimism, but would require separate validation and explicit failure conditions.

The direction of a change matters. Area-preserving relocation may have substantial risk despite passing an area heuristic, and foreground removal can have a much larger all-reference penalty than equal-sized addition. This suggests reporting removed and added counts separately in adaptation research. It also makes the envelope useful beyond parameter adaptation: a prompted model, postprocessor or ensemble can be compared with a baseline if their returned masks share the same pixel universe.

The empirical study uses one lightweight architecture and two source tasks. Three seeds measure training variation but cannot establish backbone-independent adaptation gains. The building test contains only ten tiles and four validation tiles; geographical dependence and validation variability matter. ClinicDB frames can be related, and Kvasir patient/video identities are unavailable. Exact hashes exclude duplicate files/decoded arrays, not every near-duplicate or related acquisition. Conditional bootstrap intervals may therefore understate variation across patients or regions.

The low-resolution source recipe was selected for reproducible consumer-hardware execution and does not reproduce the paper's stronger multiscale result. Its source quality constrains candidate usefulness. Synthetic appearance shifts are limited models of acquisition changes, and the clean external dataset is not a prospective clinical study. Boundary F1 has a resolution-dependent pixel tolerance. The certificate protects only the stated binary Dice score relative to a baseline; it does not protect lesion detection, boundary distance, object counts, patient outcomes or fairness. In particular, returning a baseline under severe corruption can satisfy the relative bound while remaining unusable.

The code uses floating-point arithmetic with extensive verification, not formal machine certification. The algebra assumes the same native masks and smoothing used by the evaluator. Computing an envelope on downsampled masks and then returning a differently resized mask would invalidate that correspondence. The oracle-bit ablation adds weak supervision and does not show that a clinical workflow can supply a correct bit, particularly on negative frames. Model-state adaptation across different images, multiclass averaging, 3D physical distances and multiple-candidate optimization beyond the fixed grid remain separate questions.

\section{Conclusion}
The mask-pair Dice envelope turns a vague update-safety question into an exact relative-degradation limit. A baseline-contained witness reduces the calculation to three observable counts and at most four cardinalities. A finite interpolation grid can certify partial changes without additional network forwards. The accompanying experiments measure both the guarantee and the adaptation opportunity it sacrifices. This distinction provides a reusable evaluation tool: bounded change, absolute segmentation quality and realized adaptation benefit are different properties and should be reported separately.

\paragraph{Reproducibility and assistance.} The accompanying release contains source code, six checkpoints, protocols, hashes, epoch logs, per-case candidate/decision records, analytic tests, tables and plotting scripts. Dataset files are obtained from original providers and excluded from redistribution. The source and complete execution records are supplied in the accompanying reproducibility archive. AI tools assisted implementation, mathematical development and initial drafting; all reported experimental measurements arise from executed code and preserved evidence.

\appendix
\section{Allocation algorithm and numerical edge cases}
The sharper routine computes $r=a-c$. Identical masks return zero. If $b<a$, evaluate Eq.~\eqref{eq:scalar} at $t=r$ only. Otherwise evaluate $t=r,a$ and the floor/ceiling of any in-interval stationary point from Eq.~\eqref{eq:stationary}. Return the maximum, clipped below at zero. Equal square roots require only endpoints. The positive smoothing constant defines all denominators, including empty predictions and references. Input feasibility still requires $a+b-c\leq N$ even though $N$ is absent from the optimum.

The independent benchmark routine instead forms $(r,c,o,q)$ and greedily fills all four regions. On each of up to four segments it evaluates cumulative endpoints and stationary-point neighbors of its rational objective, requiring at most thirteen sizes. Both routines are exact. Retaining the general routine in the running experiment avoids changing execution hashes or relabeling existing latency measurements after the sharper proof was found.

Counting the binary masks is $O(N)$; either ensuing routine is $O(1)$. Reference-mask reconstruction is unnecessary for gating, but all removed foreground and the requisite number of common foreground pixels identify a witness if desired. The released tests enumerate references independently of both routines. Observed-data assertions also recompute each envelope from saved counts and verify the scored decrement; rounded publication tables are not used for these checks.

\section{Complete metric and computation tables}
Table~\ref{tab:metrics} adds mean IoU and boundary F1 for the primary policies. The complete CSV files provide every condition, seed-averaged comparison, budget, sensitivity candidate, and conditional interval. These tables expose clean-data outcomes as well as corrupted-input outcomes, rather than reporting only favorable shifts.

\begin{table}[h]
\centering\scriptsize
\caption{Primary-policy secondary metrics, averaged over the same mixtures as Table~\ref{tab:main}. Native mean IoU and two-pixel-tolerance boundary F1 are percentages. The Dice certificate does not guarantee either endpoint.}
\label{tab:metrics}
\begin{tabular}{llrrr}
\toprule Cohort & Policy & IoU & Boundary F1 & Worst Dice drop\\\midrule
ClinicDB & Frozen & 41.43 & 29.16 & 0.000 \\
 & Full update & 43.55 & 30.83 & 52.571 \\
 & JS/area proxy & 41.84 & 29.55 & 17.668 \\
 & Exact rollback & 41.43 & 29.19 & 0.740 \\
 & Loose rollback & 41.43 & 29.17 & 0.458 \\
 & Metric rollback & 41.43 & 29.16 & 0.182 \\
 & Exact interpolation & 41.42 & 29.23 & 0.887 \\
 & Loose interpolation & 41.43 & 29.19 & 0.477 \\
\midrule
Buildings & Frozen & 24.61 & 34.63 & 0.000 \\
 & Full update & 25.63 & 35.84 & 0.599 \\
 & JS/area proxy & 24.65 & 34.66 & 0.552 \\
 & Exact rollback & 24.62 & 34.62 & 0.196 \\
 & Loose rollback & 24.61 & 34.62 & 0.146 \\
 & Metric rollback & 24.61 & 34.63 & 0.054 \\
 & Exact interpolation & 24.65 & 34.65 & 0.253 \\
 & Loose interpolation & 24.63 & 34.63 & 0.146 \\
\midrule
Kvasir-SEG & Frozen & 23.89 & 8.21 & 0.000 \\
 & Full update & 25.77 & 8.83 & 49.081 \\
 & JS/area proxy & 24.21 & 8.34 & 14.122 \\
 & Exact rollback & 23.89 & 8.22 & 0.786 \\
 & Loose rollback & 23.89 & 8.22 & 0.414 \\
 & Metric rollback & 23.89 & 8.22 & 0.307 \\
 & Exact interpolation & 23.92 & 8.25 & 0.864 \\
 & Loose interpolation & 23.90 & 8.23 & 0.447 \\
\bottomrule
\end{tabular}
\end{table}

\begin{table}[h]
\centering\scriptsize
\caption{Synchronized operation timings. Candidate rows include one frozen forward and the candidate/reset operations. Exact-gated rows additionally include native-mask interpolation, counting and certificate arithmetic. Blend rows reuse the primary candidate and add grid selection; labeled scoring and input reading are excluded. Forward/backward counts are means because some reference updates skip pixels.}
\label{tab:timing}
\begin{tabular}{llrrrr}
\toprule Cohort & Method & Median ms & P95 ms & Forward & Backward\\\midrule
ClinicDB & Frozen & 9.39 & 14.47 & 1.00 & 0.00 \\
 & Flip average & 19.16 & 27.51 & 2.00 & 0.00 \\
 & Full update & 52.02 & 68.80 & 3.00 & 1.00 \\
 & Exact rollback & 52.33 & 69.17 & 3.00 & 1.00 \\
 & Exact interpolation & 53.10 & 70.37 & 3.00 & 1.00 \\
 & Tent-seg & 52.13 & 69.43 & 3.00 & 1.00 \\
 & EATA-style & 104.35 & 136.79 & 5.00 & 3.00 \\
 & SAR-style & 84.06 & 109.80 & 4.00 & 2.00 \\
Buildings & Frozen & 11.47 & 19.44 & 1.00 & 0.00 \\
 & Flip average & 22.90 & 35.73 & 2.00 & 0.00 \\
 & Full update & 62.95 & 85.54 & 3.00 & 1.00 \\
 & Exact rollback & 63.59 & 86.30 & 3.00 & 1.00 \\
 & Exact interpolation & 65.23 & 88.00 & 3.00 & 1.00 \\
 & Tent-seg & 61.88 & 86.58 & 3.00 & 1.00 \\
 & EATA-style & 123.05 & 165.40 & 5.00 & 3.00 \\
 & SAR-style & 103.37 & 143.84 & 4.00 & 2.00 \\
Kvasir-SEG & Frozen & 10.21 & 16.86 & 1.00 & 0.00 \\
 & Flip average & 20.68 & 30.51 & 2.00 & 0.00 \\
 & Full update & 55.99 & 73.71 & 3.00 & 1.00 \\
 & Exact rollback & 56.39 & 74.17 & 3.00 & 1.00 \\
 & Exact interpolation & 57.29 & 75.06 & 3.00 & 1.00 \\
 & Tent-seg & 55.80 & 73.57 & 3.00 & 1.00 \\
 & EATA-style & 112.90 & 143.44 & 5.00 & 3.00 \\
 & SAR-style & 91.32 & 117.58 & 4.00 & 2.00 \\
\bottomrule
\end{tabular}
\end{table}

\section{Source learning and study provenance}
\begin{figure}[h]
\centering\includegraphics[width=\linewidth]{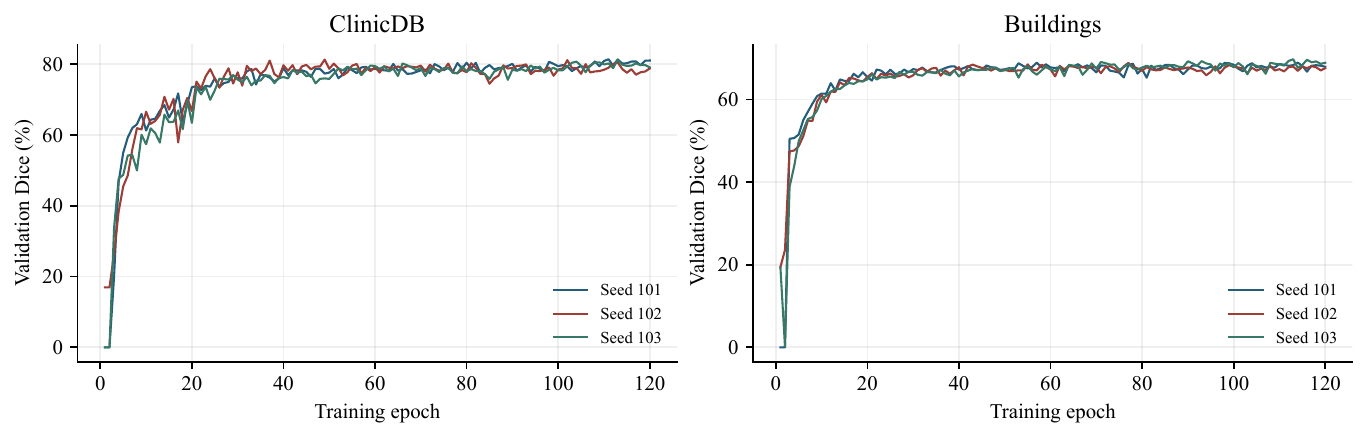}
\caption{All six 120-epoch source-training validation trajectories. Checkpoint selection uses the maximum native raw-Dice validation score within each run, not the test cohort. These are the new one-scale sources, not the earlier screening models.}
\end{figure}
The source architecture and every selected checkpoint have recorded hashes. The protocol distinguishes the original screening pilot from the present sources and external evaluation. Logs preserve all epochs, including non-best checkpoints. Image/mask alignment, red-channel building labels, native evaluation and exact decoded-image separation are audited. The mathematical simplification changes neither candidate decisions nor the reported benchmark executable; a dated extension note and full equivalence audit preserve that distinction.

\section{Recommended use of the certificate}
Use the routine on the exact binary output that would be returned. Record baseline/candidate foreground areas, intersection, total pixels, smoothing constant, certificate, decision and any interpolation alpha. A relative budget should be set by the application and disclosed; the one-point budget here is an experimental illustration. Check other task endpoints separately and retain a path for expert review when absolute source quality is inadequate. The certificate can be attached to any same-grid binary mask replacement, but an absolute quality claim requires evidence about the actual reference and intended use.


\begin{thebibliography}{99}
\bibitem{tent} D. Wang, E. Shelhamer, S. Liu, B. Olshausen, and T. Darrell. Tent: Fully test-time adaptation by entropy minimization. ICLR, 2021. \href{https://arxiv.org/abs/2006.10726}{arXiv:2006.10726}.
\bibitem{eata} S. Niu et al. Efficient test-time model adaptation without forgetting. ICML, 2022. \href{https://arxiv.org/abs/2204.02610}{arXiv:2204.02610}.
\bibitem{sar} S. Niu et al. Towards stable test-time adaptation in dynamic wild world. ICLR, 2023. \href{https://arxiv.org/abs/2302.12400}{arXiv:2302.12400}.
\bibitem{cotta} Q. Wang, O. Fink, L. Van Gool, and D. Dai. Continual test-time domain adaptation. CVPR, 2022. \href{https://arxiv.org/abs/2203.13591}{arXiv:2203.13591}.
\bibitem{onthefly} J. M. J. Valanarasu et al. On-the-Fly test-time adaptation for medical image segmentation. MIDL, 2024. \href{https://proceedings.mlr.press/v227/valanarasu24a.html}{PMLR 227}.
\bibitem{hdtta} K. Jhawar and L. Wang. HD-TTA: Hypothesis-driven test-time adaptation for safer brain tumor segmentation. Preprint, 2026. \href{https://arxiv.org/abs/2602.19454}{arXiv:2602.19454}.
\bibitem{saferestore} S. Yang and J. Wang. SafeRestore: Detector-relative risk certificates for selective industrial image restoration. Preprint, 2026. \href{https://arxiv.org/abs/2609.03475}{arXiv:2609.03475}.
\bibitem{mkunet} M. M. Rahman and R. Marculescu. MK-UNet: Multi-kernel lightweight CNN for medical image segmentation. ICCV Workshops, 2025. \href{https://arxiv.org/abs/2509.18493}{arXiv:2509.18493}.
\bibitem{kvasir} D. Jha et al. Kvasir-SEG: A segmented polyp dataset. MMM, 2020. \href{https://arxiv.org/abs/1911.07069}{arXiv:1911.07069}.
\bibitem{clinicdb} J. Bernal et al. WM-DOVA maps for accurate polyp highlighting in colonoscopy: Validation vs. saliency maps from physicians. Computerized Medical Imaging and Graphics, 43:99--111, 2015. \href{https://doi.org/10.1016/j.compmedimag.2015.02.007}{doi:10.1016/j.compmedimag.2015.02.007}.
\bibitem{mnih} V. Mnih. Machine learning for aerial image labeling. PhD thesis, University of Toronto, 2013. \url{https://www.cs.toronto.edu/~vmnih/data/}.
\bibitem{jaccard} S. Kosub. A note on the triangle inequality for the Jaccard distance. Preprint, 2016. \href{https://arxiv.org/abs/1612.02696}{arXiv:1612.02696}.
\end{thebibliography}
\end{document}